\documentclass[reprint,superscriptaddress,preprintnumbers,amsmath,amssymb,aip]{revtex4-1}

\usepackage[utf8]{inputenc}
\usepackage{graphicx}
\usepackage{grffile}
\usepackage[usenames,dvipsnames]{xcolor}
\usepackage{amsmath}
\usepackage[normalem]{ulem}
\usepackage[resetlabels, labeled]{multibib}
\usepackage{appendix}
\newcites{S}{References Supplementary Materials}
\definecolor{orange}{rgb}{1,0.5,0}
\definecolor{goodgreen}{rgb}{0.1,0.5,0}
\definecolor{goodred}{rgb}{0.7,0,0}
\usepackage{lineno}
\setpagewiselinenumbers
\modulolinenumbers[5]

\usepackage[colorlinks,urlcolor=goodgreen,citecolor=blue,linkcolor=goodred]{hyperref}
\usepackage{float}
\usepackage{babel}
\graphicspath{ {images/} }

\makeatother

\let\oldepsilon\varepsilon \let\varepsilon\varepsilon \let\varepsilon\oldepsilon
\let\oldphi\phi \let\phi\varphi \let\varphi\oldphi

\begin{document}

\title{Bipolar thermoelectrical SQUIPT (BTSQUIPT)}

\newcommand{\orcid}[1]{\href{https://orcid.org/#1}{\includegraphics[width=8pt]{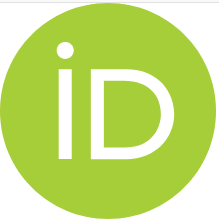}}}
\author{C. Guarcello\orcid{0000-0002-3683-2509}}
\email{Author to whom correspondence should be addressed: cguarcello@unisa.it}
\affiliation{Dipartimento di Fisica ``E.R. Caianiello'', Universit\`a di Salerno, Via Giovanni Paolo II, 132, I-84084 Fisciano (SA), Italy}
\affiliation{INFN, Sezione di Napoli Gruppo Collegato di Salerno, Complesso Universitario di Monte S. Angelo, I-80126 Napoli, Italy}
\author{R. Citro\orcid{0000-0002-3896-4759}}
\email{rocitro@unisa.it}
\affiliation{Dipartimento di Fisica ``E.R. Caianiello'', Universit\`a di Salerno, Via Giovanni Paolo II, 132, I-84084 Fisciano (SA), Italy}
\affiliation{INFN, Sezione di Napoli Gruppo Collegato di Salerno, Complesso Universitario di Monte S. Angelo, I-80126 Napoli, Italy}
\affiliation{CNR-SPIN c/o Universit\'a degli Studi di Salerno, I-84084 Fisciano (Sa), Italy}
\author{F. Giazotto\orcid{0000-0002-1571-137X}}
\email{francesco.giazotto@sns.it}
\affiliation{NEST, Istituto Nanoscienze-CNR and Scuola Normale Superiore, Piazza San Silvestro 12, I-56127 Pisa, Italy}
\author{A. Braggio\orcid{0000-0003-2119-1160}}
\email{alessandro.braggio@nano.cnr.it}
\affiliation{NEST, Istituto Nanoscienze-CNR and Scuola Normale Superiore, Piazza San Silvestro 12, I-56127 Pisa, Italy}

\begin{abstract}
We theoretically study the quasiparticle current behaviour of a thermally-biased bipolar thermoelectrical superconducting quantum interference proximity transistor, formed by a normal metal wire embedded in a superconducting ring and tunnel-coupled to a superconducting probe. In this configuration, the superconducting gap of the wire can be modified through an applied magnetic flux. We analyse the thermoelectric response as a function of magnetic flux, at fixed temperatures, in the case of a device made of the same superconductor. We demonstrate magnetically controllable, bipolar thermoelectric behaviour and discuss optimal working conditions by looking at the thermoelectric power and other figures of merit of the device.
\end{abstract}

\maketitle

Superconductivity has a long contrasted relationship with thermoelectric (TE) effects. Initially, it was considered inert thermoelectrically~\cite{Meissner} until was recognized the peculiar signature of TE effect~\cite{Ginzburg,Guttman97,Petrashov}. However hybrid and unconventional superconducting systems further extend the relevance of the linear thermoelectricity, playing with the combination of superconductors with other materials. Notable examples are superconducting-ferromagnetic systems~\cite{Machon:2013,Ozaeta:2014,Machon:2014,Kolenda:2016a,Kolenda:2016b,Kolenda:2017,Bergeret:2018,Heikkila:2018,Chakraborty:2018,Heikkila:2019,Heidrich:2019,Strambini:2022,Geng:2022} or tunnel junctions with non-conventional (iron-based) superconductors~\cite{Guarcello:2023}. In these cases, however, the particle-hole (PH) symmetry is broken extrinsically, such as in the first case due to the presence of spin polarization, or even intrinsically, such as for non-conventional superconductors or multiband superconductors that present a not PH-symmetric density of states (DOS). At the same time, intriguing phase-dependent and non-local mesoscopic effects have been also reported~\cite{Titov:2008,Jacquod:2010,Kalenkov:2017,Hussein:2019,Blasi:2020,Tan:2021} showing, unexpectedly, the richness of TE effects in superconducting systems.

More recently another phenomenon has been predicted~\cite{Marchegiani:2020a,Marchegiani:2020b,Marchegiani:2020c} and experimentally demonstrated\cite{Germanese:2022,Germanese:2023}: the bipolar TE (BTE) effect in conventional superconducting tunnel junctions made by different superconductors. The bipolar thermoelectricity reported in SIS' junctions is based on the spontaneous breaking of PH symmetry induced by a strong non-equilibrium condition, i.e., a large temperature difference associated with the peculiar features of the DOS in superconductors. In this case, the emerging thermoelectricity, which is associated to a TE power, $-IV>0$, and an absolute negative conductance (ANC), $I/V < 0$, appears necessarily \emph{bipolar}, since it can be shown that the IV characteristic is necessarily reciprocal $I(-V)=I(V)$, due to the spectral PH symmetry of the superconducting leads. Indeed, two opposite TE voltages/currents can be induced for the \emph{same} thermal gradient, giving an exclusive functionality to the BTE devices~\cite{Marchegiani:2020a}. 
\begin{figure}[t]
\includegraphics[width=0.8\columnwidth]{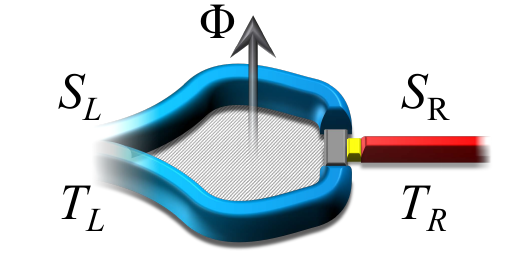}
\caption{Schematic of the device: left electrode is a superconducting loop (blue) interrupted by a small proximitized region (gray), which is tunnel-coupled through an insulating barrier (yellow) to a superconducting probe (red). Left, $S_L$, and right, $S_R$, sides reside at different temperatures, $T_L$ and $T_R$, respectively, and the loop is pierced by a magnetic flux $\Phi$.}\label{Fig01}
\end{figure}
Early experiments~\cite{Germanese:2022,Germanese:2023} exploited a normal-superconducting (NS) bilayer for adjusting the superconducting gap asymmetry, despite the inverse proximity effect affects also the sharpness of the DOS, and partially the BTE performance~\cite{Germanese:2023,Hijano:2023}. \\
Here, we propose an \emph{ad-hoc} geometry for controlling the BTE properties through an applied external magnetic field, overcoming some experimental difficulties and providing an original phase-controllable functionality in the superconducting technology. In fact, in the original proposal one cannot change the superconducting gap in a given sample, while this setup allows to fine-tune it through an external magnetic flux, thereby making it possible to adapt the BTE performance to specific operating conditions. Notably the flux dependence of this SQUIPT is quite different from the phase-dependence reported in Ref.~\onlinecite{Germanese:2023}. 
Our strategy complements other possible approaches\cite{Bernazzani:2022}, answering to different technological requirements, further widening the domain of applicability of the proposed effects for quantum technologies, energy harvesting, and energy management applications.

The device under investigation is a superconducting quantum interference proximity transistor (SQUIPT)~\cite{Giazotto:2010} formed by a normal metal wire embedded in a superconducting ring and tunnel coupled to a superconducting probe, see Fig.~\ref{Fig01}. To reveal TE effects, we need a thermal gradient across the interferometer, as recently experimentally achieved in a SQUIPT with a normal metal probe~\cite{Ligato:2022} and, crucially, the probe must be superconducting~\cite{Giazotto:2010,Ronzani2014}. Thus, in the following we assume to keep both the ring and the proximitized wire at a low temperature, $T_L\equiv T_{cold}$, and the superconducting probing electrode at high temperature, $T_R\equiv T_{hot}$. 

The DOS of the proximitized normal-metal wire can be computed using Usadel equations in a simplified 1D model~\cite{Heikkila:2002}. This reads
\begin{eqnarray}\nonumber
&\!\!\!\mathcal{N}_N(x,\varepsilon,T,\varphi)\!=\!\textup{Re}\!\Big [ \alpha(\varepsilon,T,\varphi)\cosh\!\left\{\! \frac{2x}{L} \text{arcosh}\left [ \beta(\varepsilon,T,\varphi)\! \right ]\! \right\}\!\! \Big ]\!
\end{eqnarray}
where
\begin{eqnarray}
\alpha(\varepsilon,T,\varphi)&\!=\!&\sqrt{\frac{\left ( \varepsilon+i\Gamma \right )^2}{\left ( \varepsilon+i\Gamma \right )^2-\Delta^2(T)\cos^2(\varphi/2)}},\\
\beta(\varepsilon,T,\varphi)&\!=\!&\sqrt{\frac{\left ( \varepsilon+i\Gamma \right )^2-\Delta^2(T)\cos^2(\varphi/2) }{\left ( \varepsilon+i\Gamma \right )^2-\Delta^2(T)}},
\end{eqnarray}
$\varepsilon$ is the energy relative to the chemical potential in the superconducting ring, $T$ is the wire temperature. We also took into account the phenomenological Dynes parameter $\Gamma=\gamma\Delta_0$~\cite{Dynes:1978} associated to finite quasiparticle (qp) lifetime, which was important to concretely fit SQUIPT experimental data~\cite{Ligato:2017}. In the paper, we consider a quite good-quality junction, with $\gamma_L =\gamma_R =\gamma = 10^{-4}$, but we checked that the results are only weakly (logarithmically) dependent at least for $\gamma<10^{-2}$ (data not shown).
Notably, the DOS depends on the position along the wire, given by the spatial coordinate $x\in[-L/2,L/2]$. In the case of a negligible ring inductance~\cite{Clarke:2004,Guarcello:2017}, the phase difference acquired in the superconducting ring $\varphi=2\pi\Phi/\Phi_0$ directly depends on the external magnetic flux through the loop $\Phi$, with $\Phi_0$ being the flux quantum. 
In the center of the wire, i.e., at $x=0$, the DOS reduces to 
\begin{equation}
\label{eq:DOSwire}
 \mathcal{N}_N(\varepsilon,T,\Phi)\!=\!\textup{Re}\!\left [\!\sqrt{\frac{\left ( \varepsilon+i\Gamma \right )^2}{\left ( \varepsilon+i\Gamma \right )^2- \Delta^2(T)\cos^2(\pi\frac{\Phi}{\Phi_0}) }}\right ]\!\!,
 \end{equation}
 which corresponds exactly to a standard BCS DOS where the phase-dependent gap, $|\Delta(T)\cos(\pi\Phi/\Phi_0)|$, is determined by the external magnetic flux. Therefore, the DOS can be adjusted by the magnetic flux and, consequently, we will explore how this feature reflects over the BTE performances. In this way, our bipolar TE SQUIPT (BT-SQUIPT) overcomes some of the difficulties inherent in BTE superconducting devices realized so far~\cite{Germanese:2022,Germanese:2023}, giving the possibility to explore different gap asymmetry without changing the device, thus requiring a less strict fine-tuning of temperature difference applied to the junction. 
This allows BTE effects to be observed in a device realized with a single type of superconductor with critical temperature $T_c$. This computation can be easily generalized also to the case of a probe done with a different superconductor.

In the tunneling limit the qp current flowing through the tunnel junction is simply obtained by integrating the $x$-dependent contribution of the wire over the width $w$ of the probe contact. One gets~\cite{Giazotto:2011} 
\begin{eqnarray}\nonumber
 I_{qp}=
 \frac{G_T}{ew}
 \int_{-\frac{w}{2}}^{\frac{w}{2}}dx \int d\varepsilon&& \mathcal{N}_N(x,\varepsilon,T_L,\Phi)\mathcal{N}_S(\varepsilon-eV,T_R)\\
 &&\times\mathcal{F}(\varepsilon,V,T_R,T_L),
\end{eqnarray}
where $G_T$ is the junction normal-state conductance, $\mathcal{F}(\varepsilon,V,T_R,T_L)=f_0(\varepsilon-eV,T_R)-f_0(\varepsilon,T_L)$, and $f_0(\varepsilon,T)=(e^{\varepsilon/k_B T}+1)^{-1}$ is the Fermi-Dirac distribution function since we assume that both the wire and the probe have a well defined electronic temperature. 
The DOS of the superconducting probe $\mathcal{N}_S(\varepsilon,T_R)$ is practically given by Eq.~\eqref{eq:DOSwire} for $\Phi=0$, i.e., the gap is now independent of the external magnetic flux.
In the case of a narrow superconducting probe ($w\ll L$) in contact with the center of the proximitized wire, the qp current reduces to the standard tunneling expression, where the DOS for the wire is reported in Eq.\eqref{eq:DOSwire}~\cite{Thinkambook}.

In this discussion we are completely neglecting any contribution to charge transport stemming from a dissipationless current, i.e., Josephson coupling. 
In other words, since the tunneling resistance is very high (due to the small area associated with the probe tunnel junction), the critical supercurrent $I_c$ is expected to be negligible and correspondingly the Josephson energy $E_J=\Phi_0 I_c/2\pi$ will be much smaller than the typical system operating temperature, i.e., $E_J\ll k_B T$. 
However, we have demonstrated both theoretically\cite{Marchegiani:2020c} and experimentally\cite{Germanese:2023} that a small Josephson current, although it may show a peculiar behavior~\cite{Guarcello:2019,Guarcello:2022}, does not hinder the observation of the BTE effect.

\begin{figure*}[!t]
\includegraphics[width=2\columnwidth]{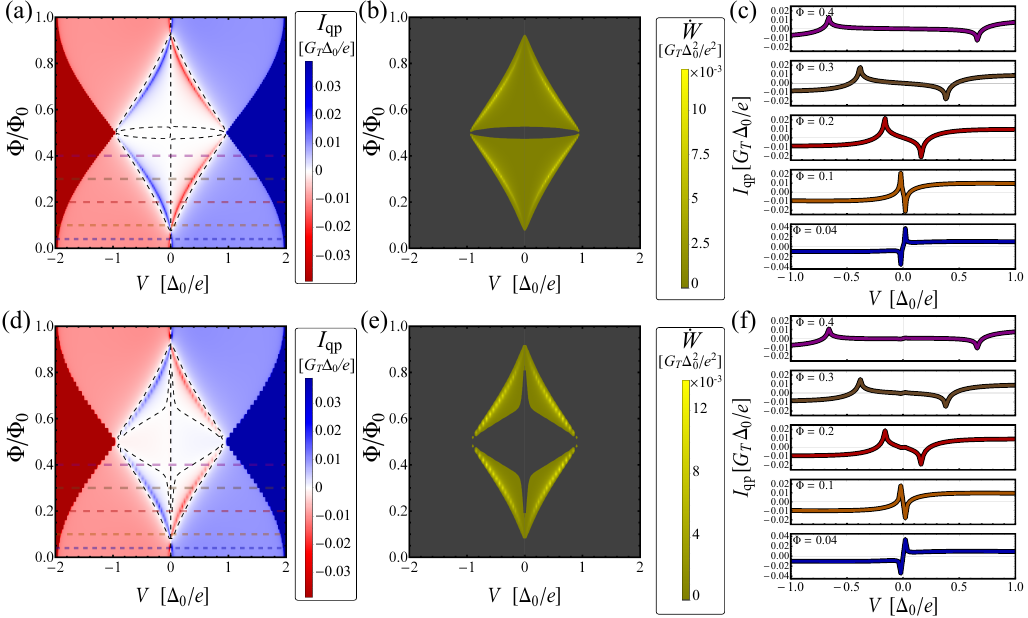}
\caption{(a-d) $I_{qp}(V,\Phi)$, (b-e) TE power $\dot{W}(V,\Phi)$, and (c-f) some selected $I_{qp}(V)$ profiles at the $\Phi$ values marked by horizontal dashed lines in (a-d), in the case of a narrow probe contact, $w\ll L$, (top panels) and $w=L/2$ (bottom panels).}\label{Fig02}
\end{figure*}

\begin{figure}[!t]
\includegraphics[width=1\columnwidth]{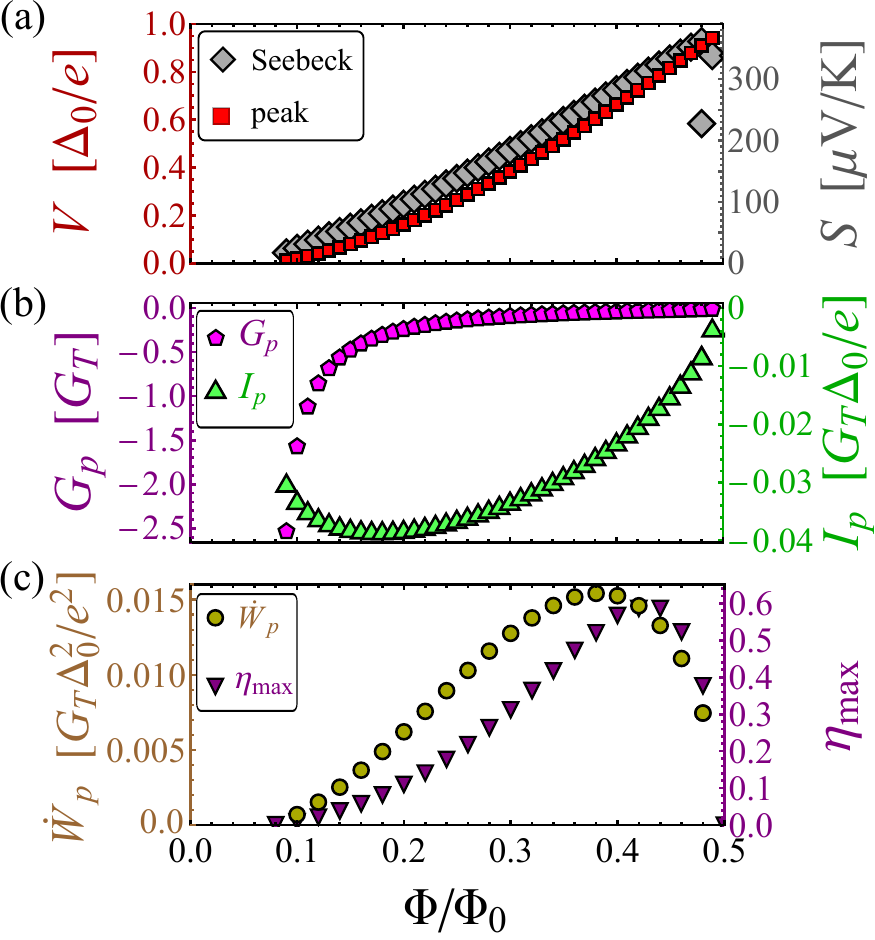}
\caption{Figures of merit of the BT-SQUIPT: (a) Seebeck and peak voltages (left vertical axis) and corresponding Seebeck coefficients (right vertical axis), (b) current, $I_p$, and conductance, $G_p$ at the matching peak, and (c) maximum TE power, $\dot{W}_p$, and efficiency, $\eta_{max}$, as a function of the magnetic flux, $\Phi$.}\label{Fig03}
\end{figure}

A SQUIPT with a superconducting probe in equilibrium has been investigated in the previous literature\cite{Giazotto:2010,Ronzani2014}.
However, since we wish to generate a bipolar thermoelectricity, one needs to apply a finite thermal gradient across the tunnel junction \cite{Marchegiani:2020a}, which we fix to $T_R=0.4 T_c$ and $T_L=0.01 T_c$ in order to guarantee a favorable out-of-equilibrium condition. 
We choose the temperature setting with the hotter probe since it is certainly the preferred condition from the experimental point of view: as a matter of fact, it is more advantageous to keep the ring and wire, which have a total volume much larger than the probe, cold, since the electron-phonon coupling scales with the volume\cite{Giazotto:2006}. 
We note that the phenomenology described below, although strongly dependent on the hot temperature $T_R$, remains qualitatively unchanged if $T_L$ changes, as long as a sufficient finite temperature difference is kept in the junction and $T_R\gtrsim0.2T_c$. 
Hereafter, we took the zero-temperature superconducting gap $\Delta_{0}$ as the standard unit of energy, and the temperature, the magnetic flux, the voltages, the qp current, and the thermopower will be given in units of $T_c=\Delta_{0}/(1.764 k_B)$, $\Phi_0$, $\Delta_0/e$, $G_T\Delta_0/e$, and $G_T\Delta^2_0/e^2$, respectively.

In the top panels of Fig.~\ref{Fig02}, we first discuss the TE response in the case of a narrow probe contact, i.e., $w\ll L$.
In Fig.~\ref{Fig02}(a) we report a density plot with the BTE current $I_{qp}(V,\Phi)$. First, we immediately observe the full odd symmetry with respect to the $V=0$ axis, that it is consequence of the reciprocity of the $IV$ characteristics, due to the PH symmetry of the DOS. At the same time, the cosine magnetic-flux-dependence of the proximitized gap, $\Delta_L(T_L,\Phi)=\Delta(T_L)\left | \cos \left ( \pi\Phi\right ) \right |$, makes the plot symmetrical with respect to $\Phi=0.5$ and periodic (in the plot we show only a single period), as required by the flux dependence of a SQUIPT. 
We firstly show the existence of the BTE effect, $\dot{W}=-I_{qp}V>0$, at subgap energies $eV< \Delta_R(T_R)+\Delta_L(T_L,\Phi)$. We also see the emergence of a full dissipative behavior $IV>0$ for $eV\gg \Delta_R(T_R)+\Delta_L(T_L,\Phi)$, which determines the strong blue/red area in the border of the density plot in Fig.~\ref{Fig02}(a). 
A dissipative response also emerges at low magnetic fluxes, i.e., for $\Phi\lesssim0.1$: in fact, there is a threshold value $\Phi_{th}^l(T_L,T_R)= \arccos{\left [\pm\Delta_R(T_R)/\Delta_L(T_L) \right ]}/\pi$, corresponding to the condition whereby the hot-side superconducting gap becomes greater than the cold-side one, below which the BTE effect no longer emerge.
This behavior will imply that there will be $(V,\Phi)$ values where $I_{qp}(V,\Phi)=0$, which are marked by black dashed lines in the density plot. 

Figure~\ref{Fig02}(b) displays the thermoelectric power $\dot{W}=-I_{qp}(V,\Phi) V$, which shows, in the yellow region, where the bipolar thermoelectricity is generated in the plane $(V,\Phi)$.

Figure~\ref{Fig02}(c) highlights a few representative $I_{qp}(V)$ profiles traced at the $\Phi\in[0-0.5]$ values marked by horizontal dashed lines in Fig.~\ref{Fig02}(a). 
As expected, due to $IV$ reciprocity, all curves cross the abscissa at $V = 0$. Furthermore, we observe low-voltage ANC and, at high-voltage, one finds Ohmic behaviour $I_{qp}\sim G_T V$ (not shown). It is easy to identify two finite voltages values $\pm V_S$ where $I_{qp}(\pm V_S)= 0$, which define the voltage bias range where the BTE effect occurs. These $V_S$ values are named \emph{Seebeck voltages} and indicate the condition in which the system is no longer able to push electrical current against the bias. The black dashed lines in the density plot of Fig.~\ref{Fig02}(a) serve to mark exactly the Seebeck voltages and to clearly identify them in the $(V,\Phi)$ parameter space. 
Thus, the region of $(V,\Phi)$ values where the system generate BTE effect (thermo-active response) is the one delimited by the black dashed line in Fig.~\ref{Fig02}(a) at $V\equiv V_S$.

The curves in Fig.~\ref{Fig02}(c) are also characterized by the presence of the peaks at $eV_p= \pm[\Delta_R(T_R)- \Delta_L(T_L,\Phi)]$, due to the matching of the BCS singularities. 
Since for $|\Phi|\leq 1/2$ the superconducting gap $\Delta_L(\Phi)$ decreases with the magnetic flux, when $|\Phi|\to1/2$ the matching peak position, $|V_p|$, increases, shifting towards higher absolute voltages. 
However, for $\Phi=1/2$ the bipolar thermoelectricity is strongly suppressed since the wire DOS becomes practically energy independent (such as in the normal metal case). In this case, the DOS looses the monotonously decreasing profile which typically characterizes BCS superconductors, and the system \emph{cannot} develop any BTE effect~\cite{Marchegiani:2020a}. 
Notably, the matching peak voltages, $|V_p|$, are always slightly smaller than the Seebeck voltages, i.e., $|V_p|\lesssim |V_S|$. The external magnetic flux can be used to tune the Seebeck voltage, even without changing the superconductor temperatures.

We note that for $\Phi\approx 1/2$, where the superconducting gap in the proximitized wire is strongly suppressed, we can face a situation already noted in Ref.~\onlinecite{Marchegiani:2020b} in which the junction at the matching peak is in the TE regime, despite the absence of ANC ($IV$ slope at zero voltages). 

In the bottom panels of Fig.~\ref{Fig02}, we report the TE response in the case of a probe contact with a finite length, $w/L=0.5$. The density plot in Fig.~\ref{Fig02}(d) shows the same qualitative behavior as observed in Fig.~\ref{Fig02}(a); the main difference we note is a larger region of $(V,\Phi)$-parameter space around $\Phi=0.5$ in which the system is dissipative. This result becomes even more evident by looking at the $\dot{W}(V,\Phi)$ map in Fig.~\ref{Fig02}(e): we can immediately see how the thermo-active region (indicated in yellow) shrinks considerably compared to Fig.~\ref{Fig02}(b). We also note how the maximum $\dot{W}$ value is well below that shown in Fig.~\ref{Fig02}(b) for $w=0$. In fact, the qp current at the matching peak significantly reduces by increasing the size of the probe contact, as can be readily seen by comparing the $I_{qp}(V)$ profiles at given magnetic fluxes shown in Figs.~\ref{Fig02}(c) and (f).

\begin{figure}[!t]
\includegraphics[width=1\columnwidth]{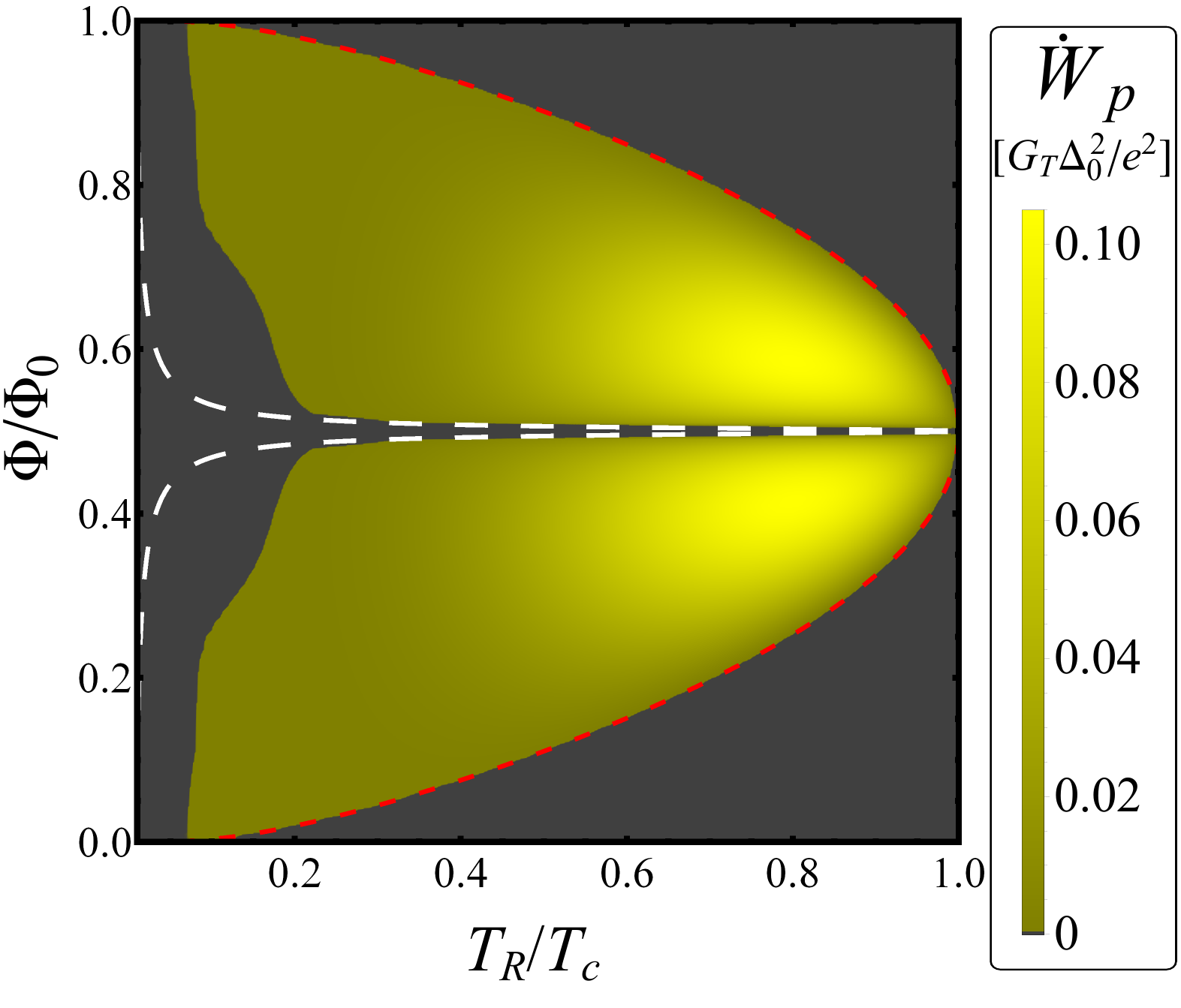}
\caption{Maximum TE power, $\dot{W}_p=-V_pI_p$, as a function of $T_R$ and $\Phi$, at $T_L=0.01$. The red and white dashed curves mark the thresholds $\Phi_{th}^l(T_L,T_R)$ and $\Phi_{th}^h(T_L,T_R)$, respectively.}\label{Fig04}
\end{figure}

In Fig.~\ref{Fig03}, we collect different figures of merit of the BT-SQUIPT as a function of the magnetic flux through the ring, calculated at $V>0$. Here we concentrate only on the case with a very narrow probe ($w\ll L$) that gives the best performance. These plots make evident the absence of BTE effect for $\Phi\lesssim 0.1$, because in such a case the gap asymmetry is not enough, with respect to the temperature difference, to generate thermoelectricity, violating the approximate BTE condition, i.e., $T_R/T_L\gtrsim\Delta_R(T_R)/\Delta_L(T_L,\Phi)$~\cite{Marchegiani:2020a}. 
Actually, this condition sets a further threshold in the magnetic flux for the TE effect to be observed, i.e., $\Phi\lesssim\Phi_{th}^h(T_L,T_R)$, with $\Phi_{th}^h(T_L,T_R)= \arccos{\left [ \pm\frac{ T_L/\Delta_L(T_L)}{T_R/\Delta_R(T_R)} \right ] }\Big/ \pi$.
Panel (a) shows both the Seebeck and matching peak voltages, $V_S$ and $V_p$, respectively (see left-vertical axis). Both increase for $\Phi\to1/2$, but $V_p$ is systematically somewhat smaller than $V_S$, as anticipated before. Clearly, when the magnetic flux approaches $\Phi=1/2$ the BTE effect sharply disappears. However the exact ``position'' where this happens is non-universal and is influenced by the other parameters, such as temperatures and Dynes parameter. On the right-vertical axis we indicate the nonlinear Seebeck coefficients, $S_i=V_i/\Delta T$, where $V_i=\{V_S,V_p\}$, which reaches the value $\sim390\mu V/K$. Notably, $S_p$ practically corresponds to the nonlinear Seebeck coefficient at the maximum power, that for this setup is almost identical to the conventional nonlinear Seebeck coefficient $S_S$. 

Figure~\ref{Fig03}(b) contains the qp current at the peak voltage, $I_p(\Phi)=I_{qp}(V_p(\Phi),\Phi)$, which is non-monotonic function with a minimum around $\Phi\sim0.18$. Clearly, the qp current is negative since it corresponds to the themoactive case with ANC. Indeed, also the absolute conductance $G_p(\Phi)=I_p(\Phi)/V_p(\Phi)$ is negative, but monotonously increasing until the BTE effect completely disappears. The latter quantity has a certain relevance from an application point of view, since it indicates the maximum load conductance that can be used by a BTE engine~\cite{Marchegiani:2020b,Germanese:2022}. Finally, Fig.~\ref{Fig03}(c) shows the magnetic-flux dependence of the maximum TE power, $\dot{W}_p(\Phi)=-I_p(\Phi) V_p(\Phi)$, and the maximum TE efficiency, $\eta_{max}(\Phi)=\text{max}_{_V} \eta(V,\Phi)$ where $\eta(V,\Phi)=\dot{W}(V,\Phi)/\dot{Q}(V,\Phi)$, with $\dot{Q}(V,\Phi)$ being the quasiparticle heat current flowing out of the electrode~\cite{Thinkambook,Giazotto:2006}. We obtain in both cases non-monotonic behaviors, with maxima $\dot{W}_p(\Phi=0.38)\sim0.016$ and $\eta_{max}(\Phi=0.42)\sim0.6$. In the case of an Al-based junction (with $\Delta_0/e\sim200\;\mu\text{V}$) with $G_T=(10\;\text{k}\Omega)^{-1}$, the maximum TE power corresponds to $\dot{W}_p\sim0.064\;\text{pW}$.

Interestingly, the maximum-thermopower condition can be further optimized by changing the hot temperature of the superconducting probe, $T_R\in(T_L,1)$, still keeping the cold side in good contact with the thermal bath at $T_L=0.01$. This is demonstrated in Fig.~\ref{Fig04}, which illustrates the behavior of $\dot{W}_p(T_R,\Phi)$. The absence of significant BTE is evident for $T_R\lesssim0.2$; furthermore, the range of magnetic fluxes giving TE effect reduces by increasing $T_R$. In fact, the thermoactive region of the $(\Phi,T_R)$-parameter space is delineated by the thresholds $\Phi_{th}^l(T_L,T_R)$ and $\Phi_{th}^h(T_L,T_R)$ defined above, which are marked by the red and white dashed curves, respectively. Finally, for $(T_R,\Phi)=(0.8,0.41)$ we achieve the maximum TE power $\dot{W}_p\sim 0.105$, which corresponds to $\sim 0.4\;\text{pW}$ for an Al-based BT-SQUIPT.

In summary, we have theoretically investigated the behaviour of the quasiparticle current in a superconducting quantum interference proximity transistor with a superconducting probe, under a sizable thermal bias. This geometry, which includes a proximitized wire enclosed in a superconducting ring pierced by a magnetic flux, allows the superconducting gap to be finely tuned. We have discussed the emergence of a bipolar thermoelectric response, at given temperatures, as a function of magnetic flux and the optimal operating conditions, considering some figures of merit and looking at both the thermoelectric power and efficiency of the device.

The possibility of controlling the superconducting gap, and thereby the thermoelectric properties, via the magnetic field makes it possible to relax the rather stringent requirements for fine-tuning of temperatures and to use a combination of different superconductors.
In fact, once a temperature difference has been settled, it is sufficient to adjust the magnetic flux to properly set the operating point. In this way, we can search for the optimal working conditions that give, for example, a more pronounced thermoactive response, simply by changing the magnetic flux, and this makes the versatility of the designed thermoelectric interferometer very attractive.

\begin{acknowledgments} F.G. and A.B. acknowledge EU’s Horizon 2020 Research and Innovation Framework Programme under Grant No. 964398 (SUPERGATE), No. 101057977 (SPECTRUM) and F.G. acknowledge PNRR MUR project PE0000023-NQSTI. A.B. acknowledges the MIUR-PRIN2022 Project NEThEQS (Grant No. 2022B9P8LN), the Royal Society through the International Exchanges between the UK and Italy (Grants No. IEC R2 192166). R.C. acknowledges the project HORIZON-EIC-2022-PATHFINDERCHALLENGES-01 GA N.101115190 – IQARO.
\end{acknowledgments}

\section*{Data Availability Statement}
The data that support the findings of this study are available from the corresponding author upon reasonable request.\\





%

\end{document}